\documentclass[twocolumn,showpacs,amsmath,superscriptaddress,
amssymb,nofootinbib]{revtex4-1}
\usepackage{dcolumn}
\usepackage{bm}
\usepackage{graphicx}

\usepackage{color}

\newcommand{\vt}{\vartheta}
\newcommand{\vp}{\varphi}
\newcommand{\tp}{\tilde{\phi}}
\newcommand{\tens}[1]{{\boldsymbol{#1}}}

\newcommand{\pa}{\partial}

\newcommand{\BM}[1]{{\mbox{\boldmath $#1$}}}

\newcommand{\be}{\begin{equation}}
\newcommand{\ee}{\end{equation}}
\newcommand{\ba}{\begin{eqnarray}}
\newcommand{\ea}{\end{eqnarray}}
\newcommand{\hh}{\, ,\hspace{0.5cm}}
\newcommand{\hhh}{\, ,\hspace{0.2cm}}

\newcommand{\eq}[1]{(\ref{#1})}

\newcommand{\n}[1]{\label{#1}}

\newcommand{\MC}[1]{{\mathcal{#1}}}
\newcommand{\hor}{\stackrel{ {\mbox{\tiny H}}}{=} }
\newcommand{\CAL}{\mathcal}

\begin{document}

\title{"Hybrid" Black Holes}
\author{Valeri P. Frolov}
\affiliation{Department of Physics, University of Alberta, Edmonton,
Alberta, Canada T6G 2E1}
\email{vfrolov@ualberta.ca}
\author{Andrei V. Frolov}
\affiliation{Department of Physics, Simon Fraser University,
8888 University Drive, Burnaby, BC, Canada}
\email{frolov@sfu.ca}

\date{\today}

\begin{abstract}
We discuss a solution of the Einstein equations, obtained by gluing the external Kerr metric and the internal Weyl metric, describing an axisymmetric static vacuum distorted black hole. These metrics are glued at the null surfaces representing their horizons. For this purpose we use the formalism of massive thin null shells. The corresponding solution is called a "hybrid" black hole. The massive null shell has an angular momentum which is the origin of the rotation of the external Kerr spacetime. At the same time, the shell distorts the geometry inside the horizon. The inner geometry of the "hybrid" black hole coincides with the geometry of the interior of a non-rotating Weyl-distorted black hole. Properties of the "hybrid" black holes are briefly discussed.
\end{abstract}

\pacs{04.70.s, 04.70.Bw, 04.20.Jb \hfill  Alberta-Thy-23-14}


\maketitle

\section{Introduction}

In this paper we present a solution of the Einstein equations for what we call a {\em "hybrid" black hole}. This solution is obtained by proper gluing of the external Kerr metric and the internal Weyl metric. These metrics are glued along the joint horizon surface, where a lightlike thin  massive shell is located. The Weyl metric describes an interior of a non-rotating black hole distorted by the presence of the shell. At the same time the shell has an angular momentum which is the origin of the `dragging into the rotation' of the external spacetime, described the Kerr metric. The event horizon is a special null limit of rigidly rotating ZAMO surfaces in the Kerr geometry, that were discussed in the recent paper by the authors \cite{FroFro}.

To glue the Kerr and Weyl metrics along a common horizon we used a general approach by Barrabes and Israel presented in their remarkable paper \cite{BI}. We shall describe this formalism adopted to our problem later. Here we just make a few remarks which might be useful for better understanding of this approach. We consider two different regions, one located outside the horizon and the other inside it, as two distinct spacetime manifolds $\MC{M}_-$ and $\MC{M}_+$. We call them Kerr and Weyl domains, respectively. Their metrics are ${g}^-_{\mu\nu}$ and ${g}^+_{\mu\nu}$, $(\mu,\nu=0,1,2,3)$, and we denote the independent coordinates in these domains by $x_-^{\mu}$ and $x_+^{\mu}$. The manifolds $\MC{M}_-$ and $\MC{M}_+$ have boundaries $\Sigma_-$ and $\Sigma_+$, respectively. They are null surfaces with respect to the corresponding metrics in the domains. We obtain a single spacetime manifold $\MC{M}=\MC{M}_-\cup\MC{M}_+$ by gluing   $\MC{M}_-$ and $\MC{M}_+$ along their boundaries, that is by making the natural identification $\Sigma_-=\Sigma_+=\Sigma$.
The Kerr (Weyl) domain $\MC{M}_-$ ($\MC{M}_+$) lies to the past (future) with respect to the common horizon $\Sigma$.
Denote by $y^a$, $a=1,2,3$, the internal coordinates on $\Sigma$. The induced metrics on $\Sigma_{\pm}$,
\be
h^{\pm}_{ab}={\pa x_{\pm}^{\mu}\over \pa y^a}{\pa x_{\pm}^{\nu}\over \pa y^b} g^{\pm}_{\mu\nu}\, ,
\ee
must be isometric. We denote this metric on $\Sigma$ by $h_{ab}$.

 Because the surface $\Sigma$ of the joint horizon is null, the three-metric ${h}_{ab}$ is degenerate. The vector $\BM{n}$ of the normal to $\Sigma$ is null, and hence it is tangent to the horizon, so that $h_{ab}n^a n^b=0$. Integral lines of this vector field are null geodesics which are the generators of $\Sigma$. In a general case, there exists an ambiguity in the choice of the parametrization of these lines, so that when one glues $\Sigma_+$ and $\Sigma_-$ one must guarantee that these parametrizations are chosen to be the same. There exist a pair of commuting Killing vectors in each of the domains, and a special linear combinations of them, with constant coefficients, are null on $\Sigma_{\pm}$. In this paper we use what is called "static" soldering, defined by identification of the corresponding advanced time coordinates over $\Sigma_-$ and $\Sigma_+$. For this choice of the soldering the parameters of the null shell at the horizon are time independent (for more details, see \cite{BI}).

The metric $g_{\mu\nu}$ is continuous at the horizon in properly chosen coordinates. However, it derivatives have jumps. The formalism developed in \cite{BI} allows one to relate these jumps to a specially chosen massive thin null shell. In this paper we demonstrate that the external (Kerr) metric can be glued with the internal (Weyl) metric along the common event horizon and calculate the parameters of the corresponding null shell.

The paper is organized as follows. In Sections~II and III we study the near horizon geometry in the Kerr and Weyl domains, respectively. In Section~IV we describe the procedure of gluing these metrics along the horizon and calculate the parameters of the massive null shell spread over the horizon. Section~V contains discussion of the obtained results and their possible generalizations.

We use units in which $G=c=1$ and the sign convention adopted in the book \cite{MTW}.

\section{Kerr domain}

\subsection{Kerr metric}

The Kerr solution depends on two arbitrary constants, mass $M$ and a rotation parameter $a=J/M\le M$, where $J$ is the angular momentum of the rotating black hole. The metric $ds_-^2$ in the Boyer-Lindquist coordinates $(t,r,\theta,\phi)$ is (see e.g., \cite{MTW,FN})
\ba\n{mds}
&&ds_-^2=d\Gamma^2+d\gamma^2\, \\
&&d\Gamma^2=A dt^2+2B dt d\phi +C d\phi^2\, ,\\
&&d\gamma^2=\Sigma\left( {dr^2\over \Delta}+ d\theta^2\right)\, ,\\
&&A=-\left(1-{2Mr\over \Sigma}\right)\, ,\
B=-{2Mra\sin^2\theta\over \Sigma}\, ,\\
&&C={P\sin^2\theta\over \Sigma}\hh P=(r^2+a^2)^2-\Delta a^2 \sin^2\theta\, ,\\
&&\Delta=r^2-2Mr+a^2\hhh \Sigma=r^2+a^2\cos^2\theta\, .
\ea
The coordinate $t$ takes values in the interval $(-\infty,\infty)$, and $\phi$ is a periodic coordinate with the period $2\pi$. The (commuting) Killing vectors for the metric \eq{mds} are
\be
\BM{\xi}_t=\partial_t\hh \BM{\xi}_{\phi}=\partial_{\phi}\, .
\ee
These vectors are uniquely specified by the following properties: (i) $\BM{\xi}_t$ is timelike and normalized to one at the infinity, and (ii) the integral lines of $\BM{\xi}_{\phi}$ are closed circles.

The Kerr metric is also invariant under the reflections
\be
(t,\phi)\to (-t,-\phi)\, , \mbox{and  } \theta\to\pi-\theta\, .
\ee

For $a<1$ the equation $\Delta=0$ has two roofs
\be
r_{\pm}=M\pm\sqrt{M^2-a^2}\, .
\ee
The larger one, $r_+$,  determines a position of the event  horizon. The angular velocity of the horizon is
\be
\Omega={a\over r_+^2+a^2}={a\over 2Mr_{+}}\, .
\ee
A linear combination of the Killing vectors (which itself is a Killing vector)
\be\n{KIL}
\BM{\eta}=\BM{\xi}_t+\Omega \BM{\xi}_{\phi}
\ee
is a null generator of the event horizon.

\subsection{Horizon metric}

The Boyer-Lindquist coordinates become singular at the horizon. Because we are interested in a near horizon geometry it is convenient to introduce so called {\em Kerr ingoing coordinates} that are regular at the (future) event horizon. For this purpose we make the following coordinate transformation
\be
dv=dt+dr_*\, ,\
d\tilde{\phi}=d\phi+a{dr\over \Delta}\, , \
dr_*=(r^2+a^2){dr\over \Delta}\, .
\ee
The Kerr metric in these coordinates $(v,r,\theta,\tilde{\phi})$  takes the form
\ba\n{KI}
ds_-^2&=&-{\Delta\over \Sigma}\omega_1^2+2 dr\, \omega_1 +\Sigma d\theta^2 +{\sin^2\theta\over \Sigma}\omega_2^2\, ,\\
\omega_1&=&dv -a\sin^2\theta d\tilde{\phi}\, ,\\
\omega_2&=& (r^2+a^2)d\tilde{\phi} - a dv\, .
\ea
The metric on the horizon can be obtained from \eq{KI} if we impose a constraint  $r=r_+$. This metric is\footnote{Here and later we use notation $\hor$ to stress that a corresponding relation is valid only on the horizon $H$.}
\be\n{MH}
dh_-^2\hor\Sigma_+ d\theta^2 +{(r_+^2 +a^2)^2 \sin^2\theta\over \Sigma_+}(d\tilde{\phi}-\Omega dv)^2\, ,
\ee
where
\be
\Sigma_+=r_+^2+a^2 \cos^2\theta\, .
\ee

It is convenient to introduce coordinates which is co-rotating with the horizon by making the transformation
\be\n{ppsi}
\psi=\tilde{\phi}-\Omega v\, .
\ee
Then the horizon metric (\ref{MH}) takes the form
\be\n{HOR}
dh_-^2= B^2( \CAL{F} d\theta^2+{\sin^2\theta\over \CAL{F}}d\psi^2)\, .
\ee
Here
\be\n{mplus}
B=\sqrt{r_+^2 +a^2}\, ,\  \CAL{F}={1+\beta^2 \cos^2\theta\over 1+\beta^2}\, ,\
\beta=a/r_+\, .
\ee
The parameter $\beta$ changes in the interval from $0$ (for a non-rotating black hole) till $1$ (for an extremely rotating one). From the form of the induced metric (\ref{HOR}) it is easy to see that the horizon area is
\be
\CAL{A}_H^-=4\pi B^2=4\pi (r_+^2 +a^2)\, .
\ee
The vector $\pa_v$ coincides with the Killing vector $\BM{\eta}$ at the horizon, and $v$ is the advanced Killing time coordinate. This Killing parametrization of the horizon generators is uniquely fixed by these conditions.

It is easy to see that the metric $dh_-^2$ is degenerate. We shall use the coordinates $y^a=(v,\theta,\psi)$ on the horizon. Here and later $a,b=1,2,3$. Let us denote $y^A=(\theta,\psi)$, $A,B=2,3$. Then one has
\be
dh_-^2\hor h_{ab}dy^a dy^b\hor h_{AB} dy^A dy^B\, .
\ee
In other words, the components of the metric $h_{ab}$ in the null direction $v$ vanish.

The surface gravity $\kappa$ at the Killing horizon of the Killing vector field $\BM{\eta}$ is defined as follows
\be\n{surfgr}
\kappa^2=-{1\over 2} \eta_{\mu;\nu}\eta^{\mu;\nu}\, .
\ee
In a general case $\kappa$ depends on the normalization of the Killing vector. For the Kerr spacetime this normalization is fixed by a requirement that $\BM{\xi}_t$ is a unit timelike vector at the infinity. The corresponding value of the surface gravity is
\be
\kappa={\sqrt{M^2-a^2}\over r_+^2+a^2}={r_+^2-a^2\over 2r_+(r_+^2+a^2)}\, .
\ee

\subsection{Null tetrads in the Kerr domain}

Let us consider a null tetrad $(\BM{k},\BM{n},\BM{m},\bar{\BM{m}})$ which in the $(v,r,\theta,\tilde{\phi})$ coordinates has components
\ba\n{tet}
k^{\mu}&=& \left( 0,-{r^2+a^2\over \Sigma},0,0\right)\, ,\nonumber\\
n^{\mu}&=& \left( 1, {\Delta\over 2(r^2+a^2)},0,{a\over r^2+a^2}\right)\, ,\\
m^{\mu}&=& {1\over \sqrt{2}(r+ia\cos\theta)}\left( ia \sin\theta,0,1,{i\over \sin\theta}\right) .\nonumber
\ea
These vectors satisfy the following orthogonality conditions
\be
(\BM{k},\BM{n})=-1\, , \  (\BM{m},\bar{\BM{m}})=1\, , \mbox{other products vanish}\, .
\ee
The directions of the vectors $\BM{k}$ and $\BM{n}$ coincide with the principle null directions of the Kerr metric.
This tetrad is regular at the future event horizon \cite{Teuk}\footnote{We change the standard notations for the real null vectors of the tetrad of the paper \cite{Teuk} $\BM{l}\to \BM{n}$ and $\BM{n}\to \BM{k}$ in order to achieve consistency with notations adopted in \cite{BI}.}. This makes it different from the Kinnersley tetrad, which is usually chosen for the separation of variable in the field equations, as well as from symmetric (Carter) null tetrad. The in-tetrad can be obtained from the latter by suitable "null rotations".

The value of the vectors of the in-tetrad (\ref{tet}) at the horizon in $(v,r,\theta,\tilde{\phi})$ coordinates is
\ba\n{KNn}
{k}^{\mu}&\hor& \left( 0,-{r_+^2+a^2\over \Sigma_+},0,0\right)\, ,\
n^{\mu}\hor \left( 1,0,0,\Omega\right)\, ,\\
m^{\mu}&\hor& {1\over \sqrt{2}(r_+ +ia\cos\theta)}\left( ia \sin\theta,0,1,{i\over \sin\theta}\right).
\ea
It is easy to see that the vector $\BM{n}$ at the horizon coincides with Killing vector $\BM{\eta}$ given by \eq{KIL}, which is the null generator of the horizon.

\subsection{Near horizon geometry}

We have two coordinate systems in the vicinity of the horizon: the ingoing Kerr coordinates $x^{\mu}=(v,r,\theta,\tilde{\phi})$ and the co-rotating ones $x_-^{\mu}=(v,r,\theta,\psi)$, where $\psi$ is defined by \eq{ppsi}. In the co-rotating coordinates the null vector $\BM{n}$ on the horizon has the following form
\be
\BM{n}\hor (1,0,0,0)\, .
\ee
We introduce two more holonomic vectors tangent to the horizon
\be
\BM{e}_2=(0,0,1,0)\hh \BM{e}_3=(0,0,0,1)\, .
\ee
In the coordinates $y_-^i=(v,\theta,\psi)$  on the surface of the horizon one has
\be\n{Khol}
\BM{e}_1=\BM{n}\hor \pa_v\hhh\BM{e}_2\hor\pa_{\theta}\hhh\BM{e}_{3}\hor\pa_{\tp}=\pa_{\psi}\, .
\ee

We need to add to this set a new vector, which is transverse to the horizon. We denote it by $\BM{N}$. Let us notice that for the calculation of the transverse extrinsic curvature one needs only to know this vector on the horizon.
The normalization conditions
\be\n{NNN}
\BM{N}^2\hor 0\hhh (\BM{N},\BM{n})\hor -1\hhh (\BM{N},\BM{e}_{2})\hor (\BM{N},\BM{e}_{3})\hor 0\, ,
\ee
uniquely determine it.  Simple calculations give
\be
{N}_{\mu}\hor \left(-1, {a^2\sin^2\theta\over2( r^2_+ +a^2)},0,0\right)\, .
\ee
Let us emphasize that the components of $\BM{N}$ are the same in both of the coordinates $x^{\mu}$ and $x_-^{\mu}$.

A transverse extrinsic curvature $\CAL{K}_{ab}$ is defined as follows \cite{BI}
\be\n{KW}
\CAL{K}_{ab}=-N_{\mu} e_b^{\nu}\nabla_{\nu} e_a^{\mu}\hh (i,j=1,2,3)\, .
\ee
The components of this object are invariant under the coordinate transformations in the bulk space and it transforms  as a tensor under the change of the coordinates on the horizon. For the chosen holonomic basis it is symmetric, $\CAL{K}_{ab}=\CAL{K}_{ba}$.

To perform these calculations we used the program GRTensor.The calculations are performed in $(v,r,\theta,\tp)$ coordinates. However, since  $\CAL{K}_{ab}$ is invariant under the coordinate transformations in the bulk space, its value remains the same in the `co-rotating' coordinate $(v,r,\theta,\psi)$.
The result is
\ba
&&\CAL{K}^{-}_{11}={r_+^2-a^2\over 2r_+(r_+^2+a^2)}\, ,\
\CAL{K}^{-}_{12}=-{a^2\sin\theta \cos\theta\over \Sigma_+}\, ,\nonumber\\
&&\CAL{K}^{-}_{13}=-{a\sin^2\theta \CAL{A}_{13}\over 2r_+\Sigma_+^2}\, , \
\CAL{K}^{-}_{22}=-{r_+(r_+^2+a^2)\over \Sigma_+}\, ,\\
&&\CAL{K}^{-}_{23}=-{a(r_+^2+a^2)\sin\theta \cos\theta\over \Sigma_+}\, ,\nonumber\\
&&\CAL{K}^{-}_{33}=-{(r_+^2+a^2)\sin^2\theta\CAL{A}_{33}\over 2r_+\Sigma_+^3}\, .\nonumber
\ea
Where
\ba
\CAL{A}_{13}&=&r_+^2(3r_+^2+a^2)+a^2(r_+^2-a^2)\cos^2\theta\, ,\n{a13}\\
\CAL{A}_{33}&=& r_+^2(r_+^2-a^2)(2r_+^2+a^2)\nonumber\\
&+&a^2(5r_+^4+2r_+^2a^2+a^4)\cos^2\theta\\
&+&a^4(r_+^2-a^2)\cos^4\theta\, .\nonumber
\ea
Let us also notice that  we use the parameter $r_+$ in the above formulas instead of the black hole mass $M$. It is easy to check that
\be\n{KK}
\CAL{K}^{-}_{vv}=\kappa\, .
\ee

\section{Weyl domain}

\subsection{Metric of a static axisymmetric distorted black hole}

Let us consider the inner domain, $\MC{M}_+$, located to the future of the event horizon. According to our assumption this is a metric of a static vacuum axisymmetric distorted black hole. It has two commuting and orthogonal Killing vectors $\BM{\zeta}_T$ and $\BM{\zeta}_\varphi$. Both of them are spacelike inside the horizon. The vector $\BM{\zeta}_\varphi$ is singled out by the property that its integral lines are closed. The other Killing vector $\BM{\zeta}_T$ is uniquely defined, up to a normalization factor, by the condition that it is orthogonal to $\BM{\zeta}_\varphi$. We fix its normalization later.

The Weyl metric is a static axisymmetric vacuum solution of the Einstein equations \cite{Weyl}. A special sub-class of these metrics, that have a regular horizon, describes {\em distorted black holes} \cite{Chandra,Chan}. Properties of static distorted black holes were discussed, e.g., in \cite{GeHa}. This metric has a Killing horizon, where $\BM{\zeta}_T$ becomes null. It separates exterior and interior domains of the Weyl-distorted black hole. In this work we focus on the interior domain only. The properties of the interior of a distorted black hole  were studied in \cite{FrSh} (see also \cite{IB}).

We write the inner metric of Weyl-distorted static black hole in the form
\ba\n{WM}
ds_+^2&=&e^{2U}F dT^2+e^{-2U+2V}\left( -{d\rho ^2\over F}+\rho^2 d\vt^2\right)\nonumber\\
&+&e^{-2U} \rho ^2 \sin^2\vartheta  d\varphi^2\, ,
\ea
where $F={2m\over \rho }-1$,  $U=U(\rho ,\vartheta )$ and $V=V(\rho ,\vartheta )$. When $U=V=0$ the metric \eq{WM} reduces to the Schwarzschild one for the interior of a non-distorted black hole.
The metric (\ref{WM}) is invariant under the following transformation
\be\n{Winv}
U=\tilde{U}+b\hhh T=e^{-b}\tilde{T}\hhh \rho=e^{b}\tilde{\rho}\hhh m=e^{b}\tilde{m}\, .
\ee

In the general case the functions $U$ and $V$ can be found by solving the Einstein equations. The potential $U$ obeys a simple linear equation, that follows from $R_{TT}=0$ (see, e.g., \cite{FrSh}). For our purpose it is convenient  to write its solution in the form \cite{man1}
\be\n{UU}
U(\rho,\vt) =\sum_{k=1}^{\infty} \alpha_k \left({R\over m}\right)^k P_k\, ,
\ee
where $\alpha_k$ are constants and
\ba\n{PZR}
P_k&=&P_k(Z)\hh Z={(\rho -m) \cos\vt\over R }\, \\
R&=&\left[m^2\cos^2\vt-F\rho^2\right]^{1/2}\, .\n{RR}
\ea
Here $P_k(\ldots)$ are Legendre polynomials.
The quantities $R/m$ and $(\rho-m)/R$ are invariant under the transformation (\ref{Winv}). In the black hole interior $R^2$ can become null and negative. However, the combination of the power of $R$ and the Legendre polynomial $P_k$, that enters the series \eq{WM}, remains finite and well defined.
One can include in the expansion (\ref{UU}) an extra term, corresponding to $k=0$. But it is a constant. We fix an ambiguity (\ref{Winv})  putting this term equal to zero.

After the solution for $U$ is given, the function $V$ can be found by integrating the expressions, that are quadratic in partial derivatives of $U$. An elegant form of a solution for $V$ was derived in \cite{man1}. Following this work we present $V$ in the form\footnote{Instead of coordinates $(x,y)$ of this work, we use $(\rho,\vt)$, connected with them as follows $x=(\rho-m)/m$ and $y=\cos\vt$.} (see also \cite{man2})
\ba
V(\rho ,\vartheta)&=&\sum_{i=1}^{\infty}\sum_{j=1}^{\infty}  { ij \alpha_i\alpha_j\over i+j}\left({R\over m}\right)^{i+j} (P_i P_j-P_{i-1} P_{j-1})\, \nonumber\\
&-&{1\over m}\sum_{i=1}^{\infty} \alpha_i  \sum_{j=0}^{i-1} A_{ij}\left({R\over m}\right)^{j}P_j\, ,\\
A_{ij}&=&(-1)^{i+j}[\rho-m(1-\cos\vt)]+\rho -m(1+\cos\vt)\, .\nonumber
\ea

As one can see from the metric (\ref{WM}), the square of the Killing vector $\BM{\zeta}_T$ is proportional to $F$ and hence this vector becomes null at $\rho=2m$. It is easy to show that this relation determines a position of the Killing horizons. One of the Einstein equations ($R_{\rho\vt}=0$) implies that there is the following relation between the values of $U$ and $V$ on the horizon \cite{GeHa}
\be
V(2m,\vt)\hor 2U(2m,\vt)-2u_0.
\ee
Possible conical singularities at  fixed points of the Killing vector $\BM{\zeta}_{\vp}$ are absent when
\be
U|_{\vt=0,\pi}\hor u_0\hh V|_{\vt=0,\pi}\hor 0\, .
\ee
In what follows we assume that these conditions are satisfied. In particular, they imply the following restrictions on the coefficients $\alpha_i$
\be
\sum_{k=1}^{\infty} \alpha_{2k-1}=0\hh \sum_{k=1}^{\infty} \alpha_{2k}=u_0\, .
\ee
It is convenient to use the following notation
\be\n{UU0}
U=\CAL{U}+u_0\, .
\ee

Local analysis of  the Weyl solutions near a regular event horizon shows that in its vicinity $\CAL{U}$ and $V$ have the following form (for details, see \cite{FrSh,IB})
\ba\n{decom}
\CAL{U}&=&\CAL{U}_0(\vt)+(1-\rho/2m)\CAL{U}_1(\vt)+\ldots\, ,\\
V&=&2\CAL{U}_0(\vt)+(1-\rho/2m) V_1(\vt)+\ldots \, .
\ea
Here `dots' denote higher in powers of $(1-\rho/2m)$ terms. For known $\CAL{U}_0$, the functions $\CAL{U}_{i\ge 1}$ and $\CAL{V}_{i\ge 1}$, which enter \eq{decom}, are defined by recursive relations. In particular, one has
\ba
\CAL{U}_1&=&\cot\vt \ \CAL{U}_{0,\vt}+\CAL{U}_{0,\vt \vt}\, ,\\
V_1&=&\cot\vt \ \CAL{U}_{0,\vt}-\CAL{U}_{0,\vt \vt}-\CAL{U}_{0,\vt}^2\, .
\ea

Using the definition (\ref{surfgr}) for the Killing vector $\zeta_T=\partial_T$ one obtains the following value of the surface gravity at the horizon
\be\n{surfm}
\hat{\kappa}_+={e^{2u_0}\over 4m}\, .
\ee
In what follows we shall sew the outer (Kerr) and the inner (Weyl) metric.
The Killing vector $\BM{\eta}$ and the corresponding Killing parametrization of the horizon of the Kerr metric are uniquely defined. However, the Killing vector $\zeta_T$ of the Weyl metric can differ on the horizon from $\BM{\eta}$ by a normalization factor $\lambda$. For this reason we define new time coordinate $t_+$ and new Killing vector $\BM{\xi}_+$ as follows
 \be
t_+=\lambda T\hh \BM{\xi}_+=\partial_{t_+}=\lambda^{-1}\partial_T\, .
\ee
The surface gravity for $\BM{\xi}_+$ is
\be\n{kap}
\kappa_+=\lambda^{-1}{e^{2u_0}\over 4m}\, .
\ee
We fix the parameter $\lambda$ later.

\subsection{Regular at the future horizon coordinates}

The Weyl metric (\ref{WM}) has a coordinate singularity on the horizon.
Let us construct new coordinates, that are regular at the horizon. Using \eq{UU0} we rewrite the metric (\ref{WM}) as follows
\ba\n{WMM}
ds_+^2&=&e^{-2u_0}\left[ e^{2\CAL{U}}F d\tilde{T}^2+e^{-2\CAL{U}+2V}\left( -{d\rho ^2\over F} +\rho^2 d\vt^2\right)\right.  \nonumber\\
&+&e^{-2\CAL{U}} \rho ^2 \sin^2\vartheta  d\varphi^2\bigg] \, ,
\ea
where $\tilde{T}=\exp(2u_0)T$. Let us introduce ingoing null coordinate (advanced time) $\tilde{v}$
\be
d\tilde{v}=d\tilde{T}-{d\rho\over F}\, .
\ee
One has
\ba\n{VVV}
&&e^{2\CAL{U}}F d\tilde{T}^2-e^{-2\CAL{U}+2V} {d\rho ^2\over F}=\\
&&e^{2\CAL{U}}F d\tilde{v}^2+2e^{2\CAL{U}} d\tilde{v}\, d\rho +4m e^{2\CAL{U}_0}[(2\CAL{U}_1-V_1)+\ldots] d\rho^2\, .\nonumber
\ea
The dots here denote linear and higher order in $(2m-\rho)$ terms. Relations (\ref{WMM}) and (\ref{VVV}) explicitly demonstrate that the Weyl-distorted metric in $(\tilde{v},\rho,\vt,\vp)$ coordinates (\ref{WMM}) is regular at the horizon.

Putting  $\rho=2m$ one obtains a  metric on the black hole surface
\be\n{HOR_m}
dh_+^2\hor 4m^2 e^{-2u_0}\left[e^{2\CAL{U}_0} d\vt^2 + e^{-2\CAL{U}_0}\sin^2\vt \ d\vp^2\right]\, .
\ee
As expected, this metric is degenerate and its components in $\hat{v}$-direction vanish.

The surface area of the distorted black hole is
\be\n{m_min}
\CAL{A}_H^+=16\pi m^2 e^{-2u_0}\, .
\ee

We shall also use another advanced time coordinate $\hat{v}$ with slightly different normalization
\be
\hat{v}=\exp(-2u_0)\tilde{v}\hh d\hat{v}=dT-e^{-2u_0}{d\rho\over F}\, .
\ee
If one  has a vector $\BM{A}$, which has components $(A^T,A^{\rho},\ldots)$ in $(T,\rho)$ coordinates, then its components in $(\hat{v},\rho)$ coordinates are $(A^{\hat{v}},A^{\rho},\ldots)$, where
\be\n{VT}
A^{\hat{v}}=A^T-e^{-2u_0}{A^{\rho}\over F}\, .
\ee
Let us also notice that the Killing vector  $\BM{\zeta}_T$ in the ingoing coordinates $\hat{x}^{\mu}=(\hat{v},\rho,\vt,\vp)$ takes the form $\partial_{\hat{v}}$.

\subsection{Null tetrads}

The sectors $({T},\rho)$ and $(\vt,\vp)$ in the Weys domain are orthogonal to each other. This implies a convenient prescription for a choice of a null tetrad. The vectors $(\hat{\BM{N}},\hat{\BM{n}},\BM{E}_2,\BM{E}_3)$ of such a tetrad have the following contravariant components in $(T,\rho,\vt,\vp)$ coordinates
\ba\n{Nne}
\hat{\BM{N}}&=&\left( -{e^{-U}\over \sqrt{2F}}, -e^{U-V}\sqrt{F}/\sqrt{2},0,0\right) \, ,\nonumber\\
\hat{\BM{n}}&=&\left( {e^{-U}\over \sqrt{2F}}, -e^{U-V}\sqrt{F}/\sqrt{2},0,0\right) \, ,\\
\BM{E}_2&=&\left(0,0,  {e^{U-V}\over \rho} ,0\right)\, ,\
\BM{E}_3=\left(0,0, 0, {e^{U-V}\over \rho \sin\vt},0\right)\, .\nonumber
\ea
These vectors obey the relations
\be
(\hat{\BM{N}},\hat{\BM{n}})=-1\hh (\BM{E}_2,\BM{E}_2)=(\BM{E}_3,\BM{E}_3)=1\, ,
\ee
the other scalar products vanish. The signs in the definition of the null vectors $\hat{\BM{N}}$ and $\hat{\BM{n}}$ are chosen so, that both of them are future directed. These vectors are chosen in the symmetric form. However they are not regular at the horizon. Regular at the horizon vectors can be obtained by a boost transformation
\be
\BM{N}_+={\sqrt{2}\over \sqrt{F}} e^{-U} \hat{\BM{N}}\hh
\BM{n}_+={\sqrt{F} e^{U}\over \sqrt{2}} \hat{\BM{n}}\, .
\ee
Using the relation
\be
2\CAL{U}-V=(2\CAL{U}_1-V_1)(1-\rho/2m)+\ldots\, .
\ee
it is easy to check that the vectors
\ba
\BM{N}_+&=&\left( -{e^{-2U}\over F},-e^{-V},0,0\right)\, ,\\
\BM{n}_+&=&\left({1\over 2},-{1\over 2}e^{2U-V} F,0,0\right)\, ,
\ea
are regular at the horizon.  In $(v_+,\rho)$ coordinates their values on the horizon are
\ba\n{Nnhor}
\BM{N}_+&\hor& \left( 2m e^{-2(\CAL{U}_0+u_0)}(2\CAL{U}_1-V_1),-e^{-2\CAL{U}_0},0,0\right)\, ,\\
\BM{n}_+&\hor& \left(1,0,0,0\right)\, .
\ea
The limit of the other two vectors of the null tetrad on the horizon is
\be\n{eehor}
\BM{E}_2\hor\left(0,0,{e^{-\CAL{U}_0+u_0}\over 2m},0\right)\, ,\
\BM{E}_3\hor\left(0,0,0,{e^{-\CAL{U}_0+u_0}\over 2m\sin\vt}\right)\, .
\ee

\subsection{Near horizon geometry}

We can use the coordinates $\hat{x}^{\mu}=(\hat{v},\rho,\vt,\vp)$ in the vicinity of the Weyl horizon.
However, as we explained earlier, there is no guarantee that the coordinate $\hat{v}$ on the horizon coincides with the Kerr advanced time $v$. So we write
\be
v=\lambda \hat{v}\hh \BM{n}=\lambda^{-1}\BM{n}_+\hh \BM{N}=\lambda \BM{N}_+\, .
\ee
For the proper choice of a constant $\lambda$ the advanced time $v$ and vectors $\BM{n}$ and $\BM{N}$ are continues at the common horizon surface. We specify $\lambda$ later and keep it in the formulas for a while.
Thus the coordinates  on the surface of the horizon $\Sigma_+$ are $y_+^i=(v,\vt,\vp)$. We introduce the following holonomic basis of vectors tangent to the horizon $\Sigma_+$
\be\n{Whol}
\BM{n}=\pa_{v}\hh\BM{e}_{\vt}=\pa_{\vt}\hh \BM{e}_{\vp}=\pa_{\vp}\, ,
\ee
One can see that vectors $\BM{e}_{\vt}$ and $\BM{e}_{\vp}$ are colinear with $\BM{E}_2$ and $\BM{E}_3$ and differ from them only by the normalization.

A transverse extrinsic curvature $\CAL{K}_{ab}$ is defined as follows \cite{BI}
\be\n{KW}
\CAL{K}_{ab}=-N_{\mu} e_b^{\nu}\nabla_{\nu} e_a^{\mu}\, .
\ee
We use the program GRTensor to calculate the components of this objects. The result is
\ba
\CAL{K}^{+}_{11}&=&-{1\over 2}\lambda^{-1} e^{2U-V} (2F U_{,\rho }+F_{,\rho })\, ,\nonumber\\
\CAL{K}^{+}_{12}&=&U_{,\vt}\hh \CAL{K}^{+}_{21}={1\over 2}V_{,\vt}\, ,\\
\CAL{K}^{+}_{22}&=&-\lambda\rho e^{-2U+V} \left[\rho  (V_{,\rho }-U_{,\rho })+1\right]\, ,\nonumber\\
\CAL{K}^{+}_{33}&=&\lambda \rho \sin^2\vt e^{-2U-V} \left[\rho  U_{,\rho }-1\right]\, ,\nonumber
\ea
Other components are zero.

Using local expansions (\ref{decom}) of $U$ and $V$ near the horizon one obtains the following expressions for the components of $\CAL{K}^{+}_{ab}$ on the horizon
\ba
\CAL{K}^{+}_{11}&=& \kappa_+\equiv {e^{2u0}\over 4\lambda m}\, ,\nonumber\\
\CAL{K}^{+}_{12}&=&\CAL{K}^{+}_{21}=\CAL{U}_{0,\vt}\, , \nonumber \\
\CAL{K}^{+}_{22}&=&-{1\over 2\kappa_+}\left[1+(\CAL{U}_1-V_1)\right]\\
\CAL{K}^{+}_{33}&=& -{1\over 2\kappa_+} e^{-4\CAL{U}_0}\sin^2\vt(1-\CAL{U}_1)\, .\nonumber\\
\ea

\section{Gluing solutions together}

\subsection{Induced geometry}

By comparing the induced metrics $dh_{\pm}^2$ on the horizon surface, given by \eq{HOR} and \eq{HOR_m}, one arrives at the following conclusions:
\begin{itemize}
\item Coordinates $\vt$ and $\vp$ on the Weyl horizon coincide with the coordinates $\theta$ and $\psi$ on the Kerr horizon;
\item The surface areas of the both horizons are identical if
\be\n{c1}
2me^{-u_0}=\sqrt{r_+^2+a^2} ;
\ee
\item The shapes of the both horizons are the same provided
\be\n{c2}
    \exp{(2\CAL{U}_0)}=\CAL{F}\equiv{1+\beta^2\cos^2\theta\over 1+\beta^2}\, ;
    \ee
\end{itemize}
In what follows we shall use $\theta$ and $\psi$ as common coordinates on the joint horizon. The conditions (\ref{c1}) and (\ref{c2}) guarantee that the induced geometries on the joint horizon coincide.

Using relations (\ref{RR}) one has at the horizon $\rho=2m$
\be
R\hor m|\cos\theta|\, ,\ \varepsilon\equiv \cos\theta/|\cos\theta|=\pm 1\, ,\  P_i\hor P_i(\varepsilon)=\varepsilon^i\, ,\nonumber
\ee
and the relation (\ref{UU}) takes the form
\be\n{UUUU}
U\hor\sum_{i=1}^{\infty} \alpha_i \cos^i\theta\, .
\ee
Using \eq{c2} one can write
\ba\n{uU}
U&\hor& u_0+\CAL{U}_0=u_0+{1\over 2}\ln \CAL{F}\, ,\nonumber\\
&=&u_0+{1\over 2}[\ln(1+\beta^2 \cos^2\theta)-\ln(1+\beta^2)]\, .
\ea

\begin{itemize}
\item By comparing this relation with \eq{UUUU} one concludes that
\be\n{c3}
u_0={1\over 2}\ln(1+\beta^2)\, ,
\ee
\item The relation (\ref{c3}) implies that
\be
e^{2u_0}={r_+^2+a^2\over r_+^2}={2M\over r_+}\, ,
\ee
and hence
\be
m=M={r_+^2+a^2\over 2r_+}\, .
\ee
\item If the surface gravities $\kappa$ and $\kappa_+$ coincide one has
\be\n{lam}
\lambda={2(r_+^2+a^2)\over r_+^2-a^2}\, .
\ee
\item
The holonomic bases $\BM{e}_i$ defined by \eq{Khol} and  \eq{Whol} are identical.
\end{itemize}

Using the above relations, one can rewrite $\CAL{K}^{+}_{ab}$ in terms of coordinates and parameters of the external Kerr solution
\ba
\CAL{K}^{+}_{11}&=& \kappa_+\equiv {e^{2u0}\over 4\lambda m}\, ,\
\CAL{K}^{+}_{12}=\CAL{K}^{+}_{21}=-{a^2 \sin\theta \cos\theta\over \Sigma_+}\, , \nonumber \\
\CAL{K}^{+}_{22}&=&(2\kappa_+\Sigma_+^2)^{-1}\times [r_+^2(r_+^2+2a^2)\\
&-&a^2(2r_+^2+3a^2)\cos^2\theta +2a^4\cos^4\theta]\, ,\nonumber\\
\CAL{K}^{+}_{33}&=&-{(r_+^2-a^2\cos^2\theta)\sin^2\theta\over 2\kappa_+ \Sigma_+^2}\, .\nonumber
\ea

\subsection{Potential $U$ in the Weyl domain}

Before considering the jump conditions for transverse extrinsic curvature let us show that the relation (\ref{c2}) determines the potential $U$ everywhere in the Weyl domain.
Using the Teylor decomposition of the depending on the angle $\theta$ part in the right-hand side of \eq{uU} one obtains
\be\n{alpha}
\alpha_{2k-1}=0\hh \alpha_{2k}=(-1)^{k+1}{\beta^{2k}\over 2k}\, .
\ee
It is easy to check that
\be\n{sumH}
\sum_{k=1}^{\infty} \alpha_{2k}={1\over 2}\ln(1+\beta^2)\, .
\ee
as it should be.

We use now the obtained value for the coefficients $\alpha_i$ and substitute them into \eq{UU}. The summation in the resulting series can be performed and one obtains (see Appendix)
\ba\n{UUU}
U&=&{1\over 2}\ln\left( {\MC{X}_+ \MC{X}_-\over 4}\right)\, ,\\
\MC{X}_{\pm}&=&\sqrt{X_{\pm}^2+\lambda}+X_{\pm}\, ,\\
X_{\pm}&=&1\pm i {y} Z\hhh \lambda={y}^2(Z^2-1)\hhh {y}=\beta R/ m\, ,\\
P_k&=&P_k(Z)\hh Z={(\rho -m) \cos\vt\over R }\, \\
R&=&\left[m^2\cos^2\vt-2m\rho +\rho^2\right]^{1/2}\, .
\ea
Let us notice that expression \eq{UUU} for the potential $U$ in the Weyl domain contains complex quantities. However, the function $U$ is real, as it should be.

One can check the obtained answer by substituting it in the equation $R_{TT}=0$. On the horizon $|Z|=1$ and $\lambda=0$, and the potential $U$ takes the form
\be
U\hor {1\over 2} \ln(X_+ X_-)\hor {1\over 2}\ln(1+{y}^2)\hor{1\over 2}\ln(1+\beta\cos^2\theta)\, .
\ee
Thus the boundary conditions are also satisfied.

\subsection{Massive thin null shell}

To satisfy the continuity of the Killing vectors at the horizon
in what follows we impose the condition
\be\n{pm}
\kappa_-=\kappa_+\, ,
\ee
and use relation (\ref{lam})
to fix the normalization of the advance time coordinates $v$. We require the continuity of $v$ in such a chosen parametrization.
Following \cite{BI} we denote by $\gamma_{ab}$ the jump of the transverse extrinsic curvature
\be
{1\over 2}\gamma_{ab}=[\CAL{K}_{ab}]\equiv \CAL{K}^+_{ab}-\CAL{K}^-_{ab}\, .
\ee
The condition (\ref{pm}) implies
\be\n{g11}
n^a n^b \gamma_{ab}=\gamma_{11}=0\hh \gamma_{12}=0\, .
\ee

In order to use the formalism developed in \cite{BI} we introduce an object $g_*^{ab}$ obeying the equation
\be
g_*^{ac}g_{bc}=\delta^a_b+n^a N_b\, .
\ee
For the adopted choice of the null vectors $\BM{n}$ and $\BM{N}$ the metric $g_*^{ac}$ can be chosen as the contravariant two-metric $g^{AB}$, bordered by zeros. Using (\ref{g11}) one can write the relation (31) of \cite{BI} in the form
\be
-16\pi S^{ab}=[g_*^{ac} n^b n^d+g_*^{bd} n^a n^c-n^a n^b g_*^{cd}]\gamma_{cd}\, .
\ee
We put $\epsilon=0$ in this relation since the horizon surface is null, and put $\eta=-1$, since $(\BM{n},\BM{N})=-1$.
Let us denote
\be
q^a=-{1\over 2} g_*^{ac} n^d \gamma_{cd}\hh
\CAL{K}= g_*^{cd}(\CAL{K}^+_{cd}-\CAL{K}^-_{cd})\, .
\ee
Then one has
\be\n{set}
8\pi S^{ab}=\CAL{K} n^a n^b + 2q^{(a} n^{b)}\, .
\ee

It is easy to check that
\be
q^a=q\xi_{\psi}^a \hh q=-{a \CAL{A}_{13}\over 2r_+ \Sigma_+(r_+^2+a^2)^2}\, ,
\ee
Here $\xi_{\psi}=\partial_{\psi}$ is the Killing vector, generating the rotations of the horizon, and $\CAL{A}_{13}$ is given by \eq{a13}.

Let us denote
\be
\mbox{tr}\CAL{K}^{\pm}=g^{CD}\CAL{K}^{\pm}_{CD}\, .
\ee
Then, since $g^{CD}$ is continuous at the horizon, one has
\be
\CAL{K}=\mbox{tr}\CAL{K}^{+}-\mbox{tr}\CAL{K}^{-}\, .
\ee

The calculations give
\ba
\mbox{tr}\CAL{K}^{-}&=-&{ 4r_+^4+3r_+^2 a^2+a^4
+a^2(r_+^2-a^2)\cos^2\theta\over 2r_+ (r_+^2+a^2)\Sigma_+}\, ,\nonumber\\
\mbox{tr}\CAL{K}^{+}&=-&{(r_+^2+a^2)r_+ B^+
\over (r_+^2-a^2)\Sigma_+^3}\, ,\\
B^+&=&r_+^2(2r_+^2+3 a^2)- (3r_+^2+4a^2)a^2\cos^2\theta\nonumber\\
&+&2a^4\cos^4\theta\, ,\nonumber\\
\CAL{K}&=& -{ a^2\sin^2\theta B \over 2r_+ (r_+^4-a^4)\Sigma_+^3}\, ,\\
B&=&r_+^4(15 r_+^4+18 r_+^2 a^2+7 a^4)\nonumber\\
&+&2r_+^2 a^2(r_+^4-6r_+^2 a^2-3a^4)\cos^2\theta\nonumber\\ &+&(r_+^2-a^2)^2 a^4 \cos^4\theta\, .\nonumber
\ea

Suppose $\Phi(x^{\mu})=0$ is the equation of the shell and $\alpha$ is defined by the relation
\be
\alpha^{-1}\partial_{\mu}\Phi\hor n_{\mu}\, ,
\ee
then, considered as a distribution, the stress-energy tensor of the shell is \cite{BI}
\be
T_{\Sigma}^{\mu\nu}=\alpha S^{\mu\nu} \delta(\Phi)\, .
\ee
We choose $\Phi=r_+-r$, so that $\Phi$ increases from $\MC{M}_-$ to $\MC{M}_-$. The vectors $\BM{n}$ and $\Phi_{,\mu}$ (in the Kerr domain) have components
\be
n_{\mu}\hor (0,{\Sigma_+\over r_+^2+a^2},0,0)\hh
\Phi_{,\mu}\hor (0,-1,0,0)\, ,
\ee
so that
\be
\alpha=-{r_+^2+a^2\over \Sigma_+}\, .
\ee
The function $\alpha$ is negative, as it should be for the null shells (see \cite{BI}).

\section{Discussions}

In this paper we constructed a new solution of the Einstein equations, which we called a "hybrid" black hole. The corresponding metric is obtained by gluing of the external Kerr metric and the internal Weyl-distorted metric along the common horizon . To glue these metrics we used the method of the massive thin null shells \cite{BI}. The horizon surface is null, so that the metric, induced on it is degenerate. For proper gluing of these geometries one needs to identify the null geodesics at the both horizons, that are their generators, and require that the transverse two-dimensional metrics are isometric. To exclude the re-parametrization freedom, one needs to fix the parameters along the horizon generators, before the gluing. For this purpose we use the Killing parameters. Namely, we required  the continuity of the advanced time coordinate $v$ defined in the external (Kerr) and the internal (Weyl) domain. The only left ambiguity in the choice of $v$ is connected with the ambiguity of the normalization of the corresponding Killing vector. This ambiguity is fixed in the Kerr domain by the condition that the timelike at infinity Killing vector has a unit norm their. To fix the normalization of the corresponding Killing vector in the Weyl domain we imposed the condition that the surface gravities  calculated at the horizon of  the Kerr and Weyl metrics are the same.

We calculated the stress-energy tensor of the massive thin null shell $S^{ab}$ (formula (\ref{set})). We demonstrated that it is a sum of two terms. The first one, ${1\over 8\pi}\CAL{K} n^a n^b$, is proportional to $a^2$. It describes the null fluid, propagating along the horizon. The second one, ${1\over 4\pi}q^{(a} n^{b)}$, is proportional to $a$ and it describes a  current along the horizon, carrying the angular momentum. Such a "rotating" null shell generates theangular momentum of the external (Kerr) metric. At the same time it distorts the "non-rotating" interior. The massive null shell "respects" both of the spacetime symmetries: it is stationary (time independent) and axisymmetric.

The horizon of the "hybrid" black hole is a special (null) case of a rigidly rotating ZAMO surface \cite{FroFro}. An interesting question is whether it is possible to construct a solution of the Einstein equations which is a junction of the Kerr and Weyl metrics, glued at a timelike surface od a rigidly rotating ZAMO surface.

\section*{Acknowledgments}

The authors thank the Natural Sciences and Engineering Research Council of Canada for the financial support. One of the authors (V.F.) is also grateful to the Killam Trust for its financial support.

\appendix

\section{Calculation of the Weyl potential $U$}

In this appendix we obtain a solution for the potential $U$ in the inner (Weyl) domain. Relations (\ref{uU}) and (\ref{c3}) imply
\be
U\hor {1\over 2}\ln (1+\beta^2\cos^2\theta)\, .
\ee
Using series expansion for $\ln(1+x)$
\be
\ln(1+x)=-\sum_{k=1}^{\infty} {(-1)^k x^k\over k}\, ,
\ee
one can write $U$ on the horizon in the form
\be
U\hor -\sum_{k=1}^{\infty} {(-1)^k (\beta\cos\theta)^{2k}\over 2k}\, .
\ee
The coefficients $\alpha_k$, that enter the solution (\ref{UU}), are of the form (\ref{alpha}). Thus one has
\be\n{serU}
U= -\sum_{k=1}^{\infty} {(-1)^k q^{2k}\over 2k}P_{2k}(Z)\, .
\ee
Here ${y}={\beta R\over m}$ and $R$ and $Z$ are defined by relations (\ref{PZR}).

To perform summation of the series (\ref{serU}) we start with the relation for a generating function of the Legendre polynomials. Denote
\be
f_{\pm}={1\over \sqrt{1-2sZ+s^2}}\, ,
\ee
then one has
\be
\sum_k s^k P_k(Z)=f_-\, .
\ee
Let us denote $s=i{y}$ then it is easy to check that
\be
U=-\int_0^{y} {d{y}\over {y}}\left[{1\over 2}(f_+ + f_-)-1\right]\, .
\ee
The integral can be easily calculated.
Let us denote
\be
X_{\pm}=1\pm i{y}Z\hh \lambda={y}^2(Z^2-1)\, ,
\ee
then one has
\ba\n{QQ}
U&=&{1\over 2}\left[ \mbox{arctanh} \left( {X_-\over \sqrt{X_-^2+\lambda}}\right)\right.\nonumber\\
&+&\left. \mbox{arctanh} \left( {X_+\over \sqrt{X_+^2+\lambda}}\right)+\ln(\lambda/4)\right]\, .
\ea
Here $\mbox{arctanh}(x)$ is an arc hyperbolic tangent of $x$, which can be expressed in terms of $\ln$ function
\be
\mbox{arctanh}(x)={1\over 2}\ln{1+x\over 1-x}\,.
\ee
After simplifications \eq{QQ} takes the form
\ba\n{QQQ}
U&=&{1\over 2}\ln\left({\MC{X}_{+}\MC{X}_{-}\over 4}\right)\, \\
\MC{X}_{\pm}&=&\sqrt{X_{\pm}^2+\lambda}+X_{\pm}\, .\nonumber
\ea


\begin{thebibliography}{}

\bibitem{FroFro} A. V. Frolov and V. P. Frolov, {\em Rigidly rotating ZAMO surfaces in the Kerr spacetime},
Phys. Rev. {\bf D 90} 124010 (2014); e-Print: arXiv:1408.6316.

\bibitem{BI} C. Barrab\'es and W. Israel, {\em Thin shells in general relativity and cosmology: The lightlike limit}, Phys. Rev. {\bf D 43}, 1129 (1991).

\bibitem{MTW} C.~W.~Misner, K.~S.~Thorne and J.~A.~Wheeler, {\em Gravitation},   W. H. Freeman and Co., San Francisco, (1973).

\bibitem{FN} V. P. Frolov and  I. D. Novikov, {\em Black Hole Physics Basic Concepts and New Developments},
Kluwer Academic Publishers (1998).

\bibitem{Teuk} S.~A.~Teukolsky, {\em Perturbations of a rotating black hole. I}, {\em Astrophys. Journ.} {\bf 185}, 635 (1973).


\bibitem{Weyl} H. Weyl, {\em Zum Gravitationstheorie}, Ann. Physik (Leipzig), {\bf 54}, 117 (1917).

\bibitem{Chandra} S. Chandrasekhar, {\em The Mathematical Theory of Black Holes}, Clarendon Press, Oxford, (1983).

\bibitem{Chan} S. Chandrasekhar, {\em On Weyl's solution for space-times with two commuting Killing fields},
    Proc. R. Soc. Lond. {\bf A 415}, 329 (1988).


\bibitem{GeHa} R. Geroch and J. B. Hartle, {\em Distorted black holes},   J. Math. Phys., {\bf 23}, 681 (1982).

\bibitem{FrSh} V.~P.~Frolov and A.~Shoom, {\em Interior of Distorted Black Holes}, Phys. Rev. {\bf D 76}, 064037 (2007).

\bibitem{IB} T.~Pilkington, A.~Melanson, J.~Fitzgerald and I.~Booth, {\em Trapped and marginally traped surfaces in Weyl-distorted Schwarzschild solutions},  Class.Quant.Grav. {\bf 28}, 125018 (2011).

\bibitem{man1} N. Bret\'{o}n, T. E. Denisova, V. S. Manko, {\em  Kerr black hole in the external gravitational field},  Physics Letters {\bf A 230}, 7, (1997).

\bibitem{man2} N. Bret\'{o}n, A. A. García, V. S. Manko, and T. E. Denisova, {\em Arbitrarily deformed Kerr-Newman black hole in an external gravitational field},  Phys. Rev. {\bf D 57}, 3382 (1998).




\end{thebibliography}
\end{document}